\documentclass[aps,prb,reprint,amsmath,amssymb,floatfix]{revtex4-2}
\usepackage[T1]{fontenc}
\usepackage[utf8]{inputenc}
\usepackage[bookmarks=false,linkcolor=NavyBlue,urlcolor=NavyBlue,colorlinks,citecolor=NavyBlue]{hyperref}
\usepackage{graphicx}
\usepackage{amsmath}
\usepackage[svgnames]{xcolor}
\usepackage{enumerate}
\usepackage[normalem]{ulem}
\usepackage{orcidlink}
\usepackage{comment}

\graphicspath{{./Figures/}}

\newcommand{\figref}[2]{\hyperref[#1]{\ref{#1}(#2)}}
\newcommand{\figsref}[2]{\hyperref[#1]{\ref{#1}#2}}
\newcommand{\sigmoid}{\sigma_{\rm sig}}

\begin{document}

\title{Competing nonlinearities, criticality, and order-to-chaos transition in deep networks}

\newcommand{\Cornell}{Department of Physics, Cornell University, Ithaca, NY 14853, USA}

\author{Omri~Lesser\,\orcidlink{0000-0002-8616-6284}}
\affiliation{\Cornell}

\author{Debanjan~Chowdhury\,\orcidlink{0000-0003-0758-0282}}
\affiliation{\Cornell}

\date{\today}

\begin{abstract}
Deep neural networks owe their expressive power to nonlinear activation functions.
The effective field theory of signal propagation at initialization reveals a few distinct universality classes of activations that exhibit different depth scaling. Tuning across these, especially with analytical control, is an open problem. We show that a statistical mixture of activations, where each neuron independently and randomly draws its activation from a two-component distribution with mixing fraction $p$, provides a new mechanism for a continuous phase transition. Applied to a mixture of \texttt{Tanh} and \texttt{Swish}, the transition is sharp in the depth scaling of the preactivation variance, separating a variance-collapsing from a variance-inflating phase; at $p_c$, the network acquires statistical scale invariance, with depth-independent variance, without sacrificing smoothness. This resolves a longstanding tension, where scale-invariant propagation has previously required the non-smooth \texttt{ReLU} family, rendering such networks ill-suited to curvature-based optimizers, physics-informed architectures, and neural-network quantum states. We corroborate the transition through variance propagation, parallel and perpendicular susceptibilities, and Lyapunov exponents. Training multilayer perceptrons on real datasets reveals non-monotonic test performance as a function of $p$, with an optimum near the theoretically predicted $p_c$, confirming that the initialization-level transition has direct consequences for learned representations. The quenched activation disorder acts as a structural regularizer, suppressing memorization of corrupted labels while preserving generalization. Our framework establishes statistical activation mixtures as a controlled tool for navigating the phase diagram of deep network universality classes.
\end{abstract}

\maketitle

\tableofcontents

\section{Introduction}

The capacity to train deep neural networks rests on the ability to propagate information effectively through many layers. During gradient-based training, signals traveling forward and gradients traveling backward generically grow or shrink exponentially with depth, making learning impractical~\cite{glorot_understanding_2010}. In the limit of large network width, this becomes analytically tractable: preactivations at each layer converge to a Gaussian distribution with zero mean and a variance that obeys a deterministic, layer-to-layer recursion~\cite{neal_priors_1996,lee_deep_2018, matthews_gaussian_2018}. This recursion is determined entirely by the choice of activation function and weight initialization, and constitutes the effective field theory of the network at initialization~\cite{poole_exponential_2016, schoenholz_deep_2017,bahri_statistical_2020, roberts_principles_2022,ringel_applications_2025}. The condition for stable training is \emph{criticality}, the boundary between exponential growth and exponential decay of the variance, where signals maintain their magnitude across arbitrarily many layers. Critically initialized networks sit at the edge of chaos~\cite{sompolinsky_chaos_1988}, where information propagates without distortion and gradients neither vanish nor explode~\cite{doshi_critical_2023}.

\begin{figure*}[ht]
    \centering
    \includegraphics[width=\linewidth]{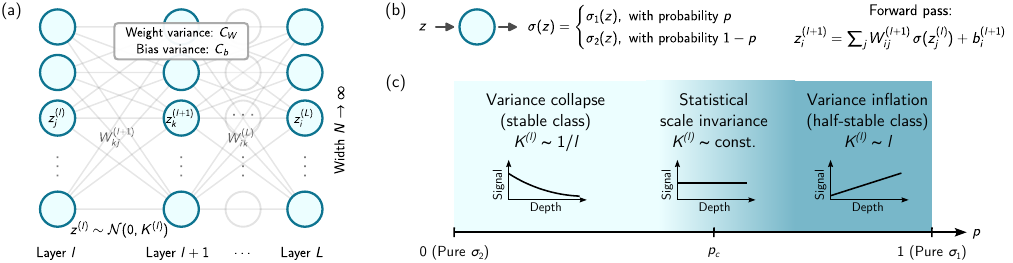}
    \caption{(a)~Schematic of the dynamics of variance propagation in a fully connected network. The preactivations $z^{(l)}$ at each layer $l$ are Gaussian distributed with zero mean and variance $K^{(l)}$. The variance evolves according to the kernel recursion Eq.~\eqref{eq:kernel_recursion}, which depends on the activation function $\sigma$ through the kernel function $g(K)$.
    (b)~The activation function is chosen randomly for each neuron.
    (c)~Schematic phase diagram: the mixing fraction $p$ controls the relative weight of two competing activations with opposing variance stability characters, with an expected phase transition at a critical $p_c$ where the network becomes statistically scale invariant.
    }
    \label{fig:network_schematic}
\end{figure*}

The requirement for criticality places sharp constraints on the nonlinear 
activation function, $\sigma$. A central result of the effective field theory is that activation functions partition into distinct \emph{universality classes}, determined entirely by the qualitative structure of the variance recursion near its fixed point $K^\star$---the value of the variance that remains invariant with depth~\cite{poole_exponential_2016, 
schoenholz_deep_2017, roberts_principles_2022}. Two properties of the fixed 
point determine the class: its location ($K^\star = 0$ or $K^\star > 0$) and 
its linear stability (whether nearby variance trajectories are attracted to or repelled from it). The rectified linear unit (\texttt{ReLU}) has a 
special position: its scale invariance, $\sigma(\alpha z) = 
\alpha\sigma(z)$, forces the kernel recursion to be exactly linear, rendering the network automatically critical at $K^\star = 0$ for \emph{any} 
initialization. This property is widely credited as a key factor in 
\texttt{ReLU}'s empirical success~\cite{glorot_deep_2011, 
poole_exponential_2016, schoenholz_deep_2017, roberts_principles_2022}, and 
has been extended to convolutional and residual architectures through the 
theory of dynamical isometry~\cite{xiao_dynamical_2018,pennington_resurrecting_2017}. However, 
this scale invariance comes at the cost of a non-smooth kink at the origin; 
\texttt{ReLU} is not differentiable at $z = 0$ and has a vanishing second 
derivative everywhere else. This makes it ill-suited to applications where 
smoothness is not merely a convenience but a requirement, including 
curvature-based and natural-gradient optimizers that rely on well-defined 
Hessians~\cite{amari_natural_1998, martens_new_2020}, physics-informed neural networks that solve partial differential equations by differentiating through the network~\cite{raissi_physics-informed_2019}, and neural-network quantum states whose variational energy involves derivatives of the 
wavefunction~\cite{carleo_solving_2017, lange_architectures_2024}.

Modern activations such as \texttt{Swish} and \texttt{GELU} were designed to 
combine the favorable propagation properties of \texttt{ReLU} with smooth, 
infinitely differentiable profiles~\cite{hendrycks_gaussian_2016, 
ramachandran_searching_2017, clevert_fast_2016}. However, their smoothness 
inevitably introduces a characteristic length scale into the problem; unlike 
\texttt{ReLU}, they are not scale-invariant, and their variance recursions 
have a qualitatively different structure. Specifically, $K^\star = 0$ is an 
\emph{unstable} fixed point for \texttt{Swish} and \texttt{GELU}---a small 
perturbation away from zero variance is amplified rather than absorbed---and 
the variance instead flows to a finite, stable fixed point $K^\star > 0$. We 
refer to this as the \emph{half-stable} class. At this finite fixed point, the network is critical in the sense that variance is depth-independent, but the fixed point itself is sensitive to initialization and introduces non-universal features that depend on the specific activation 
function~\cite{roberts_principles_2022}. Saturating activations such as 
\texttt{Tanh} and \texttt{Sin} belong to yet another class: $K^\star = 0$ is 
a \emph{stable} fixed point, so variance is attracted to zero and decays 
algebraically with depth ($K^{(l)} \sim 1/l$), leading to signal 
attenuation. The three classes---scale-invariant (\texttt{ReLU}), half-stable (\texttt{Swish}, \texttt{GELU}), and stable (\texttt{Tanh}, 
\texttt{Sin})---are thus separated by qualitative differences in long-depth 
behavior, and represent discrete labels rather than points on a continuum. 
Whether these boundaries can be crossed continuously, by tuning a single 
parameter, has not been systematically addressed to the best of our knowledge.

Here we introduce a framework for crossing universality-class boundaries using \emph{statistical mixtures} of activation functions. The central idea is to treat the activation function itself as a random variable: each neuron 
independently draws its activation from a fixed two-component distribution, 
applying $\sigma_1$ with probability $p$ and $\sigma_2$ with probability $1-p$. This is distinct from the deterministic combination $\sigma(z) = 
p\sigma_1(z) + (1-p)\sigma_2(z)$, in which every neuron applies a fixed 
weighted superposition of two functions. In the deterministic (``coherent'') case, the variance recursion contains cross-correlation terms 
$\langle\sigma_1(z)\sigma_2(z)\rangle_K$ that introduce a nonlinear 
dependence on $p$. In our statistical (``incoherent'') mixture, because each 
neuron draws one activation or the other as a mutually exclusive event, 
self-averaging in the infinite-width limit eliminates all cross terms. The 
effective kernel function becomes a strict linear interpolation between the 
pure-component kernels, and $p$ appears as an analytically transparent, 
linear control parameter. The analogy to quantum mechanics is instructive: the deterministic combination is the neural-network counterpart of a coherent superposition, whose observables contain interference contributions, while our statistical mixture corresponds to an incoherent mixed state, whose observables are weighted averages with no cross 
terms~\cite{sakurai_modern_2020, nielsen_quantum_2010}. The same 
incoherent or \emph{quenched} structure arises in the statistical physics of disordered systems~\cite{mezard_spin_1987,bahri_statistical_2020,pei_statistical_2025}, 
where quenched disorder refers to frozen heterogeneity that is fixed at 
initialization rather than resampled at each forward pass. The mixing fraction $p$ thus serves as an exact, closed-form control parameter for interpolating between universality classes.

The idea of assigning different activation functions to individual neurons has been explored empirically as an ensemble strategy~\cite{harmon_activation_2017,
maguolo_ensemble_2020}, and stochastic switching between activations has 
recently been applied in large language models to improve inference efficiency and output diversity~\cite{lomeli_stochastic_2025}. Here we provide the theoretical foundation that these works lack: a mean-field theory showing that such mixtures constitute a controlled mechanism for navigating the phase diagram of universality classes.

Mixing activations with opposing $K^\star = 0$ stability characters, 
specifically \texttt{Tanh} (stable) and \texttt{Swish} (half-stable), leads 
to a sharp phase transition at a critical probability $p_c$. We compute $p_c$ analytically in the small-variance limit and perturbatively at finite input variance, and corroborate our findings through numerical simulations of variance propagation, the parallel susceptibility $\chi_\parallel$ (which 
measures how a global rescaling of the input magnitude propagates through 
depth) and the perpendicular susceptibility $\chi_\perp$ (which measures how 
a small transverse perturbation between two nearby inputs grows or contracts 
layer by layer), and 
Lyapunov exponents. At $p_c$, the network exhibits emergent statistical scale invariance: depth-independent variance across all layers, despite being composed entirely of smooth, differentiable neurons. We further demonstrate that the transition is not merely an initialization-level phenomenon: training multilayer perceptrons on MNIST~\cite{lecun_gradient_1998} and Fashion-MNIST~\cite{xiao_fashion_2017} reveals non-monotonic test performance as a function of $p$, with an optimum near the theoretically predicted $p_c$. Finally, we show that the quenched activation disorder acts as an implicit regularizer in overparameterized networks, suppressing memorization of corrupted labels while preserving the capacity to learn genuine structure.

Before proceeding further, we note a structural analogy that places our results in a broader context. Measurement-induced phase transitions (MIPTs) in monitored quantum circuits~\cite{li_quantum_2018, skinner_measurement_2019, 
nahum_quantum_2017, fisher_random_2023} separate a 
volume-law entangled phase from an area-law phase at a critical measurement 
rate $p_c$, where entangling unitary gates compete against disentangling projective measurements, and tuning the relative frequency of each drives a transition in the long-time, large-system entanglement structure. The overarching structure of our problem is strikingly similar. In both cases, a single parameter $p$ controls the relative weight of two competing local operations with opposing tendencies: variance-inflating versus variance-collapsing in our setting, entangling versus disentangling in the quantum circuit setting. Self-averaging in the appropriate thermodynamic limit renders the phase boundary analytically tractable. In both cases the transition is continuous, diagnosed by a correlation-like quantity 
(the Lyapunov exponent here, the entanglement entropy there), and the critical point is characterized by emergent scale invariance. The analogy is not merely superficial: in both settings the transition is between a phase where information is preserved only locally with depth (area-law / variance collapse) and one where it proliferates (volume-law / variance explosion), with a scale-invariant critical point that supports robust information propagation. 

The remainder of this article is organized as follows.
In Sec.~\ref{sec:theory}, we develop the mean-field theory of statistical activation mixtures, derive a closed-form expression for the critical mixing fraction $p_c$, and characterize the transition through the stability coefficient $a_1$ that governs the approach to the fixed point. In Sec.~\ref{sec:tanh_swish}, we present numerical simulations of variance propagation, susceptibilities, and Lyapunov exponents that corroborate our theoretical predictions. In Sec.~\ref{sec:learning}, we demonstrate a potential utility of the proposed framework through learning experiments on established datasets, showing that the quenched disorder acts as a regularizer that improves generalization in overparameterized networks. We conclude in Sec.~\ref{sec:conclusion} with a discussion of implications and future directions. The appendices contain supporting numerical results and related discussion.

\section{Theoretical framework}
\label{sec:theory}

In this section, we develop the mean-field theory of statistical activation mixtures in three steps. We first review the kernel recursion formalism that governs variance propagation in the infinite-width limit, and recall how it partitions activation functions into distinct universality classes. We then introduce the statistical mixture construction and show that, in contrast to deterministic weighted combinations of activations, self-averaging renders the effective kernel linear in the mixing fraction $p$, making it an analytically transparent control parameter. Finally, we exploit this linearity to derive a closed-form expression for the critical mixing fraction $p_c$ at which the network undergoes a phase transition between universality classes, and characterize the transition through the stability coefficient $a_1$ that governs the approach to the fixed point.

\subsection{Mean-field dynamics and kernel recursion}
We consider fully connected networks, or multi-layer perceptrons (MLPs), of width $N$ and depth $L$.
In the infinite-width limit ($N \to \infty$), the 
central limit theorem guarantees that the preactivations $z^{(l)}$ at layer $l$ are governed by a Gaussian distribution with zero mean and variance $K^{(l)}$, for any choice of activation function and weight 
distribution with finite second moment~\cite{neal_priors_1996, 
lee_deep_2018, matthews_gaussian_2018, roberts_principles_2022, helias_lecture_2026}.
As depicted in Fig.~\figref{fig:network_schematic}{a--b}, the variance propagates through the layers according to the deterministic recursion map
\begin{equation}\label{eq:kernel_recursion}
    K^{(l+1)} = C_W g(K^{(l)}) + C_b,
\end{equation}
where $C_W$ and $C_b$ are the variances of the weights and biases at initialization (properly normalized), and $g(K)$ is the kernel function defined as the expected squared activation over the Gaussian measure:
\begin{equation}
    g(K) \equiv \langle \sigma^2(z) \rangle_K = \int_{-\infty}^{\infty} \frac{dz}{\sqrt{2\pi K}} e^{-\frac{z^2}{2K}} \sigma^2(z).
\end{equation}
The entire dependence on the activation function is thus encoded in $g(K)$; 
different choices of $\sigma$ produce different recursion maps, and the 
long-depth behavior of $K^{(l)}$ is determined by the fixed-point structure 
of this map.

For generic initialization, $K^{(l)}$ either grows or decays exponentially with depth, in both cases preventing effective learning. Criticality corresponds to the existence of a stable fixed point $K^{\star}$ of the recursion Eq.~\eqref{eq:kernel_recursion}, satisfying
\begin{equation}\label{eq:K_fixed_point}
    K^{\star} = C_{W} g\left( K^{\star} \right) + C_{b},
\end{equation}
at which the variance remains bounded and nonzero across all layers~\cite{poole_exponential_2016, schoenholz_deep_2017}.
Stability of the fixed point is captured by two susceptibilities. The \emph{parallel} susceptibility $\chi_{\parallel}$ measures how a small rescaling of the overall input magnitude propagates through the network, i.e., how sensitive the variance $K^{(l+1)}$ is to a change in $K^{(l)}$. The \emph{perpendicular} susceptibility $\chi_{\perp}$ measures how a small perturbation \emph{orthogonal} to the input (a displacement transverse to the overall scale direction) grows or shrinks from layer to layer; equivalently, it governs how quickly two nearby inputs diverge, and is therefore directly related to the sensitivity of the output to input 
perturbations~\cite{poole_exponential_2016, schoenholz_deep_2017}. These susceptibilities are given by
\begin{subequations}
\label{eq:susceptibilities}
\begin{align}
    \chi_{\parallel}(K) &= C_{W} g'(K) = \frac{C_{W}}{K} \left\langle z \sigma'(z) \sigma(z) \right\rangle_K, \label{eq:chi_parallel} \\
    \chi_{\perp}(K) &= C_{W} \left\langle \sigma'(z)^2 \right\rangle_K . \label{eq:chi_perp}
\end{align}
\end{subequations}

At a stable fixed point, both susceptibilities are equal to unity: $\chi_{\parallel}(K^{\star})=\chi_{\perp}(K^{\star})=1$. Intuitively, 
$\chi_\parallel = 1$ means that the overall signal scale is preserved from 
layer to layer, while $\chi_\perp = 1$ means that two distinct inputs neither converge nor diverge exponentially with depth --- a necessary condition for the network to remain sensitive to input differences across many layers~\cite{schoenholz_deep_2017}.

Two activation functions belong to the same \emph{universality class} if they share the same qualitative behavior near the fixed point: the location of $K^{\star}$, its stability, and the rate at which $K^{(l)}$ approaches it~\cite{poole_exponential_2016, roberts_principles_2022}.
Three classes are relevant here.
\texttt{ReLU} is scale-invariant [$\sigma(\alpha z)=\alpha\sigma(z)$], so its kernel function is exactly linear, $g(K) \propto K$, 
and the network is automatically critical and stable for \emph{any} initialization $K$. \texttt{Tanh} and \texttt{Sin} belong to the \emph{stable} class: $K^{\star}=0$ is a stable fixed point and the variance decays algebraically ($K^{(l)} \sim 1/l$), leading to signal attenuation in deep networks.
\texttt{Swish} and \texttt{GELU} belong to the \emph{half-stable} class: $K^{\star}=0$ is unstable, but there exists a finite stable fixed point $K^{\star} > 0$ at which the network is critical.
These classes are separated by qualitative differences in long-depth behavior, and our goal is to determine whether they can be continuously bridged by tuning a single parameter.

\subsection{Mixture of activations}
To study transitions between universality classes, we consider networks in which the activation function is itself a random variable drawn independently for each neuron from a distribution ${\cal P}(\sigma)$. When the network size goes to infinity, self-averaging implies that the kernel function $g(K)$ and the susceptibilities $\chi_{\parallel}(K)$, $\chi_{\perp}(K)$ are replaced by their averages over ${\cal P}(\sigma)$:
\begin{equation}
    g(K) = \int {\cal D}\sigma \, {\cal P}(\sigma) \langle \sigma(z)^2 \rangle_K.
\end{equation}
Self-averaging here is a consequence of the central limit theorem in the infinite-width limit: because each neuron's preactivation is a sum of $N$ independent contributions, sample-to-sample fluctuations in the empirical kernel are suppressed as $1/\sqrt{N}$ and vanish as $N \to \infty$~\cite{bahri_statistical_2020, roberts_principles_2022}. This is the same mechanism that renders the Gaussian process description of infinite-width networks exact~\cite{neal_priors_1996, lee_deep_2018}: the 
quenched disorder in the activation assignments contributes a frozen heterogeneity at the level of individual neurons, but this heterogeneity is averaged out at the level of the layer-wise variance by the law of large numbers. At finite width $N$, self-averaging is approximate and sample-to-sample fluctuations are $\mathcal{O}(1/\sqrt{N})$; our numerical simulations at $N = 500$ use multiple random seeds precisely to verify that 
these fluctuations are small and that the mean-field predictions are quantitatively accurate at this width.
As the simplest case, we consider a two-component distribution:
\begin{equation}
    {\cal P}(\sigma) = 
    \begin{cases}
        p, & \sigma = \sigma_{1} \\
        1-p, & \sigma = \sigma_{2} \\
        0, & {\rm otherwise,}
    \end{cases}
\end{equation}
with $\sigma_{1}(z)$, $\sigma_{2}(z)$ two given (deterministic) activation functions.
This is a Bernoulli mixture of activations: each neuron applies activation function $\sigma_1$ with probability $p$ and $\sigma_2$ with probability $1-p$, independently of all other neurons, and this assignment is 
fixed (quenched) for the lifetime of the network.
Due to the linearity of the expectation value, the effective kernel function for the mixture becomes a linear interpolation,
\begin{equation}\label{eq:g_mix}
    g^{(\rm mix)}(K) = p g^{(\sigma_{1})}(K) + (1-p) g^{(\sigma_{2})}(K).
\end{equation}
The same relation holds for the susceptibilities:
\begin{equation}
    \chi_{\parallel,\perp}^{(\rm mix)} (K) = p \chi_{\parallel,\perp}^{(\sigma_{1})} (K) + (1-p) \chi_{\parallel,\perp}^{(\sigma_{2})} (K).
\end{equation}
To ensure the network is initialized at criticality for $K^{\star}=0$ (unit slope at the origin), we set $C_b=0$ and choose $C_W$ as:
\begin{equation}
\begin{aligned}
    C_W &\left[g^{(\rm mix)}\right]'(0) = 1 \\
    &\implies C_W = \frac{1}{p \langle (\sigma'_1)^2 \rangle_0 + (1-p) \langle (\sigma'_2)^2 \rangle_0}.
\end{aligned}
\end{equation}

We stress that these relations are distinct from those found in the usual scenario of mixing activations~\cite{agostinelli_learning_2014,manessi_learning_2018,harmon_activation_2017}, where each neuron applies a fixed deterministic combination of two functions, i.e., $\sigma(z)=p\sigma_{1}(z) + (1-p)\sigma_{2}(z)$.
Since both $g(K)$ and $\chi_{\parallel,\perp}(K)$ involve products of $\sigma$, they contain cross terms, or ``interference'' terms, absent from our model.
It is instructive to draw an analogy to quantum mechanics~\cite{sakurai_modern_2020,nielsen_quantum_2010}.
A coherent superposition $|\psi\rangle = \sqrt{p}|\psi_1\rangle + \sqrt{1-p}|\psi_2\rangle$ produces expectation values $\langle\psi|O|\psi\rangle$ that include interference contributions $\langle\psi_1|O|\psi_2\rangle$.
An incoherent (mixed) state, described by the density matrix $\rho = p|\psi_1\rangle\langle\psi_1| + (1-p)|\psi_2\rangle\langle\psi_2|$, yields only the weighted average $\mathrm{Tr}(O\rho) = p\langle\psi_1|O|\psi_1\rangle + (1-p)\langle\psi_2|O|\psi_2\rangle$, with no cross terms.
Our statistical mixture is the neural-network analogue of the incoherent case: because each neuron draws one activation or the other as a mutually exclusive event, all observables average independently, and the linearity of 
Eqs.~\eqref{eq:susceptibilities}--\eqref{eq:g_mix} follows exactly. The same 
structure arises in the statistical physics of disordered systems with \emph{quenched} disorder~\cite{mezard_spin_1987, 
bahri_statistical_2020, pei_statistical_2025}: heterogeneity that is frozen 
at initialization rather than resampled at each forward pass. 
In the deterministic (``coherent'') scenario, an explicit cross-correlation kernel $\tilde{g}(K) \equiv \langle\sigma_1(z)\sigma_2(z)\rangle_K$ enters the variance recursion:
\begin{equation}\label{eq:g_coherent}
\begin{aligned}
g^{(\rm coh)}(K) &= p^2\,g^{(\sigma_1)}(K) + (1-p)^2\,g^{(\sigma_2)}(K) \\
&+ 2p(1-p)\,\tilde{g}(K).
\end{aligned}
\end{equation}
The cross term $\tilde{g}(K)$ can be computed perturbatively by expanding $\sigma_1$ and $\sigma_2$ in Taylor series near $K=0$.
As long as $\sigma_{1,2}(0)=0$ and $\sigma'_{1,2}(0)\neq 0$, a standard necessary and sufficient condition~\cite{roberts_principles_2022}, $K^\star=0$ remains a valid fixed point at any $p\in(0,1)$.
Stability and critical mixing, however, depend on $\tilde{g}(K)$ in a nonlinear way, so the coherent recursion does not reduce to the clean linear interpolation of pure-component kernels that characterizes our statistical mixture, and the location of $p_c$ (if it exists) cannot be written in closed form.

\subsection{Stability analysis and universality classes}
The behavior of the network near the fixed point $K^{\star}=0$ determines its universality class.
We expand the recursion Eq.~\eqref{eq:kernel_recursion} around $K=0$.
For a smooth activation, we write $\sigma(z) = \sum_n \frac{\sigma_n}{n!} z^n$, which yields  $g(K) = g_1 K + g_2 K^2 + \cdots$, with coefficients $g_n$ determined by the Taylor coefficients $\sigma_n$~\cite{roberts_principles_2022}. At criticality 
($C_W g_1 = 1$, $C_b = 0$), the leading-order deviation from the fixed point 
obeys
\begin{equation}
    \Delta K^{(l+1)} = \Delta K^{(l)} + a_1 (\Delta K^{(l)})^2 + \mathcal{O}((\Delta K^{(l)})^3),
\end{equation}
where the stability coefficient $a_1$, which controls the sign and rate of 
the algebraic approach to $K^\star = 0$, is given by~\cite{roberts_principles_2022},
\begin{equation}
    a_1 \equiv \left( \frac{\sigma_3}{\sigma_1} \right) + \frac{3}{4} \left( \frac{\sigma_2}{\sigma_1} \right)^2.
\end{equation}

For completeness, we record the explicit forms of the two activation 
functions used throughout. \texttt{Tanh} is the standard hyperbolic tangent,
\begin{equation}
    \sigma_{\rm Tanh}(z) = \tanh(z) = \frac{e^z - e^{-z}}{e^z + e^{-z}},
\end{equation}
which is an odd, bounded, saturating function with $\sigma_{\rm Tanh}(0) = 0$ 
and $\sigma'_{\rm Tanh}(0) = 1$. \texttt{Swish} is defined 
as~\cite{ramachandran_searching_2017}
\begin{equation}
    \sigma_{\rm Swish}(z) = z \cdot \sigmoid(z) = 
    \frac{z}{1 + e^{-z}},
\end{equation}
which is smooth, unbounded, and approximately linear for large $|z|$, with 
$\sigma_{\rm Swish}(0) = 0$ and $\sigma'_{\rm Swish}(0) = 1/2$. Both 
functions are \emph{parameter-free}: neither contains a tunable scalar that 
would shift the Taylor coefficients $g_n$ and consequently the value of 
$p_c$. This is a deliberate choice that keeps the analysis clean and 
$p$ as the sole control parameter. For parameterized variants (e.g. 
\texttt{Swish}-$\beta$, defined as $z \cdot \sigmoid(\beta z)$, whose $a_1$ 
depends on $\beta$), the general formula Eq.~\eqref{eq:pc_general} still 
applies, but $p_c$ acquires an additional dependence on the parameter $\beta$, tracing a 
critical curve in the $(p, \beta)$ plane rather than a critical point on the 
$p$ axis. Mapping such critical manifolds in the space of parameterized activation functions is a natural extension of the present work.

When $a_1 < 0$, the fixed point is stable and $\Delta K$ decays algebraically ($\Delta K^{(l)} \sim 1/l$); this is the $K^\star=0$ class, whose prominent examples are \texttt{Tanh} and \texttt{Sin}.
When $a_1 > 0$, the fixed point is unstable: the variance is repelled from zero and flows to a finite $K^\star \neq 0$; this is the half-stable class, which includes \texttt{Swish} and \texttt{GELU}. The sign of $a_1$ 
thus serves as the order parameter for the universality class. 
We note that \texttt{ReLU} sits precisely at the boundary $a_1 = 0$, but not through the smooth Taylor-expansion mechanism above: its scale invariance forces $g(K) \propto K$ exactly, so $g_2 = 0$ identically and the fixed point is marginal to all orders.

Since $g^{(\rm mix)}$ is linear in $p$, the Taylor coefficients inherit the same linearity: $g_n^{(\rm mix)} = p\,g_n^{(\sigma_1)} + (1-p)\,g_n^{(\sigma_2)}$. The effective stability coefficient for the mixture is therefore
\begin{equation}\label{eq:a1mix}
    a_1^{(\rm mix)}(p) = \frac{g_2^{(\rm mix)}(p)}{g_1^{(\rm mix)}(p)} = \frac{p \, g_2^{(\sigma_1)} + (1-p) \, g_2^{(\sigma_2)}}{p \, g_1^{(\sigma_1)} + (1-p) \, g_1^{(\sigma_2)}}.
\end{equation}
A phase transition occurs at the critical probability $p_c$ where $a_1^{(\rm mix)}(p_c) = 0$, which is equivalent to $g_2^{(\rm mix)}(p_c)=0$, giving
\begin{equation}\label{eq:pc_general}
    p_c = \frac{g_2^{(\sigma_2)}}{g_2^{(\sigma_2)} - g_2^{(\sigma_1)}}.
\end{equation}

This is one of the main results of our mean-field theory, namely a closed-form expression for the critical mixing fraction in terms of a single Taylor coefficient of each pure-component kernel. The transition exists whenever $g_2^{(\sigma_1)}$ and $g_2^{(\sigma_2)}$ have opposite signs, i.e., whenever $\sigma_1$ and $\sigma_2$ belong to opposing universality classes. At $p_c$, the effective stability coefficient $a_1^{(\rm mix)}$ vanishes, the power-law approach to $K^\star = 0$ is eliminated, and the network acquires the same marginal, scale-invariant behavior as \texttt{ReLU}, while remaining composed entirely of smooth neurons. The qualitative phase diagram is illustrated in Fig.~\figref{fig:network_schematic}{c}.

The above criterion for a phase transition immediately rules out mixtures involving \texttt{ReLU}, whose exact scale invariance forces $g_2^{(\rm ReLU)} = 0$ and places $p_c = 1$ outside the physically accessible range for any choice of second component (see Appendix~\ref{app:relu}). The minimal nontrivial realization therefore requires one activation from the stable class ($a_1 < 0$) and one from the half-stable class ($a_1 > 0$), so that $a_1^{(\rm mix)}(p)$ interpolates through zero at a finite $p_c \in (0,1)$. In the following section we study this transition in detail using \texttt{Tanh} as the stable component and \texttt{Swish} as the half-stable component, a pairing that admits fully analytical treatment and is representative of the broader class of stable/half-stable mixtures.
\begin{figure}
    \centering
    \includegraphics[width=\linewidth]{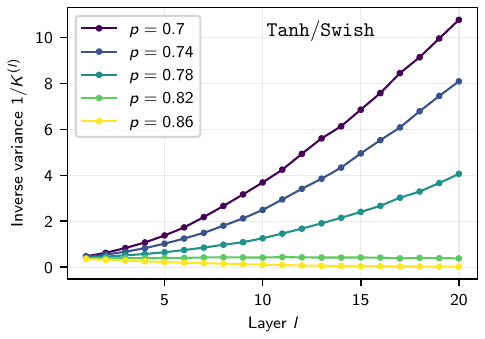}
    \caption{Inverse variance $1/K^{(l)}$ vs.\ depth $l$ for a \texttt{Tanh}/\texttt{Swish} activation mixture, for several values of the mixing fraction $p$.
    Two distinct regimes appear. For $p<p_c$ the \texttt{Tanh}-dominated network drives variance to zero ($K^{(l)}$ decays), while for $p>p_c$ the \texttt{Swish}-dominated network explodes ($K^{(l)}$ grows), both with a power-law behavior. At the empirical critical point $p_c\approx0.83$ the inverse variance is depth-independent: this is a transition between universality classes, with emergent statistical scale invariance. }
    \label{fig:variance_plots}
\end{figure}

\section{Criticality from competing fixed-point instabilities: 
the \NoCaseChange{\texttt{Tanh}/\texttt{Swish}} transition}\label{sec:tanh_swish}

Having established the general criterion for a phase transition in 
Eq.~\eqref{eq:pc_general}, we now study its consequences in the minimal 
nontrivial realization: a Bernoulli mixture of \texttt{Tanh} and 
\texttt{Swish}. These two activations are natural antagonists in the 
universality-class sense. \texttt{Tanh} belongs to the stable class: its 
saturating profile suppresses large preactivations, driving the variance 
toward zero with depth. \texttt{Swish} belongs to the half-stable class: its 
approximately linear behavior for large arguments allows the variance to grow, 
repelling the network from $K^\star = 0$ toward a finite fixed point. Neither activation is scale-invariant, so neither is automatically critical. The question is whether their competition, mediated by the mixing fraction $p$, can produce an emergent critical point at which the network behaves as if it were scale-invariant without suffering from \texttt{ReLU}'s non-smoothness. We address this question analytically, verify it numerically through three diagnostics, and confirm that the transition has measurable consequences for learning.

\begin{figure*}
    \centering
    \includegraphics[width=\linewidth]{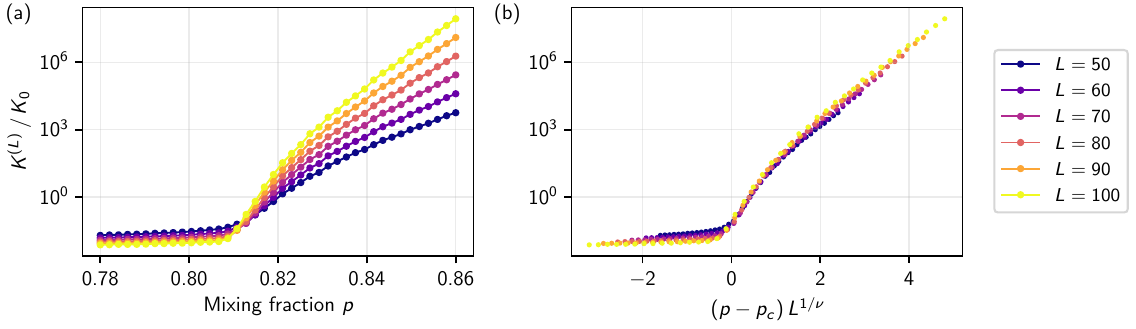}
    \caption{Finite-size scaling near the critical point.
    (a)~Variance at layer $L$ (normalized by the input variance $K_{0}$) vs.\ mixing fraction $p$ for several values of depth $L$. The variance transition sharpens with increasing depth, and the curves cross near the critical point $p_c$.
    (b)~Data collapse of the variance curves using the scaling variable $(p - p_c)\,L^{1/\nu}$. The extracted critical exponent is $\nu=1$, in agreement with a continuous mean-field-like phase transition.
    }\label{fig:fss}
\end{figure*}

\subsection{Analytical prediction of \NoCaseChange{$p_c$}}

For \texttt{Tanh} we have seen that $a_{1} = -2$, and for \texttt{Swish} it can be shown that $a_{1}=3/4$~\cite{roberts_principles_2022}.
Since the signs are opposite, there must exist a crossing point.
Using Eq.~\eqref{eq:pc_general} with $g_2^{(\rm Swish)} = 3/16$ and $g_2^{(\rm Tanh)} = -2$, we find
\begin{equation}\label{eq:pc_tanh_swish}
    p_c = \frac{g_2^{(\rm Tanh)}}{g_2^{(\rm Tanh)} - g_2^{(\rm Swish)}} = \frac{32}{35} \approx 0.91.
\end{equation}

This analysis holds in the small-variance limit: we assume the input variance $K_0$ is small enough that it can be taken as infinitesimal, so that the Taylor expansion of $g^{(\rm mix)}(K)$ around $K=0$ is controlled.
Real datasets, however, have finite input variance, and this generically shifts $p_c$ away from its mean-field value. The direction and magnitude of the shift can be computed perturbatively.

At finite input variance $K_0>0$, the network is critical when the fixed 
point $K^\star = K_0$ is simultaneously stationary ($\phi(K_0)=K_0$) and marginal ($\phi'(K_0)=1$). Together, these two conditions yield the exact criticality condition
\begin{equation}
    \frac{K_0\, \left[g^{(\rm mix)}\right]'(K_0)}{g^{(\rm mix)}(K_0)} = 1,
\end{equation}
which reduces to $g_2^{(\rm mix)} = 0$ at $K_0 = 0$, recovering 
Eq.~\eqref{eq:pc_general}. Expanding $g^{(\rm mix)}(K) = g_1^{(\rm mix)} K + g_2^{(\rm mix)} K^2 + g_3^{(\rm mix)} K^3 + \cdots$ and simplifying, this reduces to
\begin{equation}
    g_2^{(\rm mix)}(p) + 2\,g_3^{(\rm mix)}(p)\, K_0 + \mathcal{O}(K_0^2) = 0.
\end{equation}
Linearizing around $p_c^{(0)}=32/35$ where $g_2^{(\rm mix)}=0$, we obtain the corrected critical probability
\begin{equation}\label{eq:pc_corrected}
\begin{aligned}
    p_c(K_0) &= \frac{32}{35} - \frac{2\,g_3^{(\rm mix)}(p_c^{(0)})}{g_2^{(\rm Swish)} - g_2^{(\rm Tanh)}}\, K_0 + \mathcal{O}(K_0^2) \\
    &= \frac{32}{35} - \frac{384}{1225}\, K_0 + \mathcal{O}(K_0^2),
\end{aligned}
\end{equation}
where we used $g_3^{(\rm Tanh)}=17/3$, $g_3^{(\rm Swish)}=-5/32$, giving $g_3^{(\rm mix)}(32/35) = 12/35$. The negative coefficient  of $K_0$ means that finite input variance pushes $p_c$ \emph{downward} from $0.91$. A larger \texttt{Swish} fraction is needed to counteract the stronger 
variance-collapsing tendency of \texttt{Tanh} when the inputs are large. 
Note, however, that Eq.~\eqref{eq:pc_corrected} is a perturbative 
expression valid for $K_0 \ll 1$; for $K_0 = 1$ (the value used in our 
simulations), the correction term $384/1225 \approx 0.31$ is not small, and 
higher-order terms will contribute. The perturbative analysis therefore 
predicts the \emph{direction} of the shift reliably, but the precise 
numerical value of $p_c$ at $K_0 = 1$ must be determined numerically.

\subsection{Numerical diagnostics}

\subsubsection{Variance propagation}

We performed a sweep of $p$ in randomly initialized MLPs and analyzed the evolution of the inverse variance $1/K^{(l)}$ with the depth $l$, as shown in Fig.~\ref{fig:variance_plots}.
Networks of width $N = 500$ and depth $L = 20$ were used, with 20 random seeds for each value of $p$; we have verified that the qualitative picture is unchanged for deeper networks (see Appendix~\ref{app:variance_plots}).
The inputs are random Gaussian vectors with variance $K_0=1$ and dimension $D=100$.

We observe two distinct regimes, separated by a critical value $p_c \approx 0.83$. In the \texttt{Tanh}-dominated regime ($p < p_c$), 
the saturating character of \texttt{Tanh} wins: the inverse variance grows 
linearly with depth ($1/K^{(l)} \sim l$), meaning the variance collapses 
algebraically to zero. In the \texttt{Swish}-dominated regime ($p > p_c$), 
the variance-inflating character of \texttt{Swish} wins: the inverse 
variance decays with $l$, meaning the variance grows without bound. 
At $p_c \approx 0.83$, the two tendencies cancel and the variance 
profile is flat, depth-independent, and mimics the scale-invariant behavior 
of \texttt{ReLU} while being composed entirely of smooth neurons. The 
observed value $p_c \approx 0.83$ is shifted downward from the 
small-variance prediction $p_c^{(0)} \approx 0.91$, in the direction 
predicted by Eq.~\eqref{eq:pc_corrected}. We have verified that reducing 
$K_0$ pushes $p_c$ upward toward $0.91$, as expected, though this requires 
more random seeds for numerical stability due to the slower variance 
dynamics near $K = 0$; see Appendix~\ref{app:variance_plots}.

\begin{figure*}[t]
    \centering
    \includegraphics[width=\linewidth]{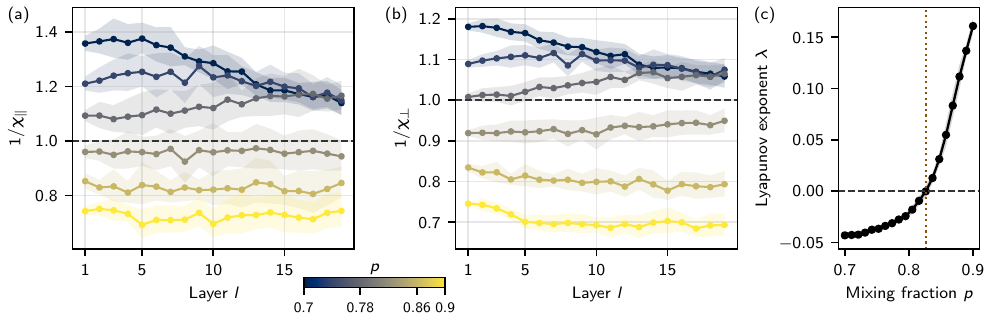}
    \caption{Signatures of criticality.
    (a)~Parallel susceptibility $\chi_{\parallel}$ and (b)~perpendicular susceptibility $\chi_{\perp}$ vs.\ layer depth $l$ for \texttt{Tanh}/\texttt{Swish} mixtures at several values of $p$. $\chi_\parallel=1$ means the overall signal scale is preserved layer-to-layer; $\chi_\perp=1$ means transverse perturbations neither grow nor shrink. Away from $p_c$, both quantities drift monotonically with depth. Near $p_c\approx0.83$, both susceptibilities remain pinned to unity across all layers, providing a signature of criticality.
    (c)~Maximal Lyapunov exponent $\lambda$ vs.\ mixing fraction $p$ for \texttt{Tanh}/\texttt{Swish} networks. $\lambda>0$ signals exponential divergence of perturbations (chaotic regime, \texttt{Swish}-dominated), while $\lambda<0$ signals contraction (ordered regime, \texttt{Tanh}-dominated). The zero-crossing at $p_c\approx0.83$ marks the critical point.}\label{fig:susceptibilities}
\end{figure*}

The sharpness of the transition with an effective system size provides an important diagnostic of a continuous phase transition.
In our setting, the role of the system size is played by the network depth $L$: it is the only 
thermodynamic-like variable that controls how many iterations of the variance recursion are applied, and therefore how far the system can evolve from its initial condition before the output is read off. In the limit $L \to \infty$, the transition between the variance-collapsing and 
variance-inflating phases becomes sharp; at finite $L$, it is rounded on a 
scale set by the correlation depth $\xi \sim |p - p_c|^{-\nu}$, the number 
of layers over which deviations from criticality accumulate appreciably.
Figure~\figref{fig:fss}{a} shows the variance $K^{(L)}$ (normalized by the input
variance $K_{0}$) as a function of $p$ for several values of depth $L$.
With increasing $L$, the transition from the decaying phase to the
exploding phase becomes progressively sharper, and the curves cross near
$p_c$, as expected from finite-size scaling near a continuous
phase transition. To quantify this, we perform a data collapse using the
scaling variable $(p - p_c)\,L^{1/\nu}$.
Figure~\figref{fig:fss}{b} shows that all curves collapse onto a single
universal branch with critical exponent $\nu = 1$, indicating that the
correlation ``length'' (here, the depth over which the variance deviates
appreciably from criticality) diverges as $|p - p_c|^{-1}$. This exponent
is consistent with a mean-field continuous transition, as expected for a
system governed by a single relevant perturbation, which is effectively the 
stability coefficient $a_1^{(\rm mix)}(p)$, which vanishes linearly at 
$p_c$ by construction, Eq.~\eqref{eq:a1mix}. This provides an 
\emph{a posteriori} justification for the mean-field treatment: the 
transition is controlled by a single relevant direction in the space of 
kernel maps, with all higher-order Taylor coefficients constituting 
irrelevant perturbations in the renormalization-group sense.

For a complementary perspective, we note that the correlation depth $\xi \sim |p - p_c|^{-\nu}$ also has an operational meaning beyond the definition as the scale over which deviations from criticality accumulate. A network of depth $L \ll \xi$ 
cannot distinguish whether it is in the collapsing or exploding phase from 
its output statistics alone, since the variance $K^{(L)}$ is still close to the 
input variance $K^{(1)}$. Such a network is effectively critical regardless of the value of $p$, which explains why shallow networks are less sensitive to the precise value of $p$ and why 
the optimal $p$ for learning in shallow networks is less sharply defined. 
Conversely, a network of depth $L \gg \xi$ is deep enough to fully 
develop the asymptotic phase: the variance has either collapsed to zero or 
grown large, and the network is far from criticality for any $p \neq p_c$. 
The correlation depth thus sets a natural lower bound on the network depth 
required to benefit from the critical initialization: networks shallower 
than $\xi(p)$ gain little from tuning $p$, while networks deeper than 
$\xi(p)$ are strongly sensitive to the distance from criticality.

\subsubsection{Susceptibilities}

A second, independent diagnostic of the transition is provided by the parallel and perpendicular susceptibilities, $\chi_{\parallel,\perp}$, as shown in Fig.~\figref{fig:susceptibilities}{a--b}. Recall that $\chi_\parallel = 1$ means the overall signal scale is preserved layer-to-layer, while $\chi_\perp = 1$ means that two nearby inputs neither decay nor diverge with depth. Both conditions must hold simultaneously at a critical point. Away from $p_c$, both susceptibilities drift monotonically with depth: $\chi_{\parallel,\perp} < 1$ in the \texttt{Tanh}-dominated phase (signals contract) and $\chi_{\parallel,\perp} > 1$ in the \texttt{Swish}-dominated phase (signals expand). 
We find that near $p_{c}\approx0.83$, both susceptibilities are approximately equal to unity and are independent of the layer $l$, providing a sharp and independent confirmation of criticality.

Both susceptibilities were estimated via finite differences on the forward pass.
To estimate $\chi_\parallel$, we scale the base preactivation $z^{(l)}$ by a 
factor $(1 + \varepsilon)$ and track the resulting fractional change in the 
empirical variance $K^{(l)} = \frac{1}{N}\|z^{(l)}\|^2$,
\begin{equation}
    \hat{\chi}_\parallel \approx \left\langle
        \frac{K^{(l+1)}_{\rm scaled} - K^{(l+1)}}{K^{(l)}_{\rm scaled} - K^{(l)}}
    \right\rangle_{\rm batch}.
\end{equation}
To estimate $\chi_\perp$, we instead pass a base preactivation $z^{(l)}$ and a perturbed version $z^{(l)} + \delta z^{(l)}$ through the layer, where $\delta z^{(l)}$ is a small random vector orthogonalized against $z^{(l)}$ (so that $\delta z^{(l)} \cdot z^{(l)} = 0$) to isolate the transverse direction and measure the ratio of output to input perturbation norms:
\begin{equation}
    \hat{\chi}_\perp \approx \left\langle
        \frac{\|z^{(l+1)}_{\rm perturbed} - z^{(l+1)}\|^2}{\|\delta z^{(l)}\|^2}
    \right\rangle_{\rm batch}.
\end{equation}
With automatic differentiation, both susceptibilities can be computed without finite differences by using Jacobian-vector products, which avoid floating-point instability of finite differences at small $\varepsilon$: $\chi_\perp$ corresponds to the mean squared singular value of the layer Jacobian $\partial z^{(l+1)}/\partial z^{(l)}$, while $\chi_\parallel$ is its derivative with respect to the input variance evaluated at the fixed point $K^{\star}$. We have verified that both methods yield identical results.

\begin{figure*}
    \centering
    \includegraphics[width=\linewidth]{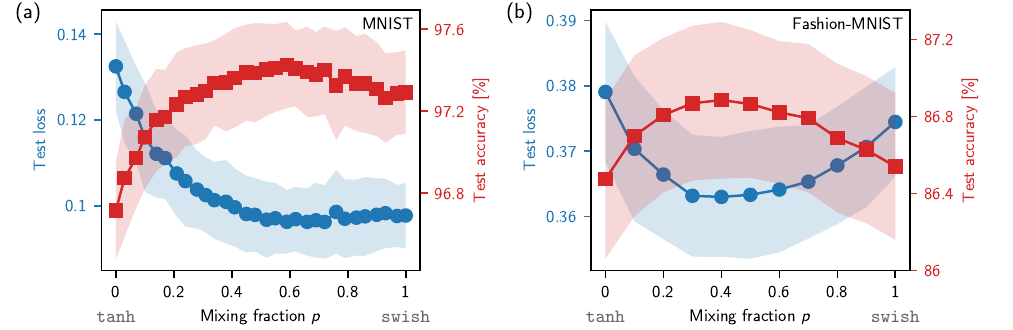}
    \caption{Performance of an MLP with a mixture of \texttt{Tanh} and \texttt{Swish} activations on (a)~the MNIST digit classification task~\cite{lecun_gradient_1998} and (b)~the Fashion-MNIST fashion items classification task~\cite{xiao_fashion_2017}.
    Both test loss (blue) and test accuracy (red) vary non-monotonically with the mixing fraction $p$, exhibiting an optimum at an intermediate $p_c$. Pure \texttt{Tanh} ($p=0$) and pure \texttt{Swish} ($p=1$) both underperform.
    The shaded regions indicate the standard deviation across 10 random seeds.
    }
    \label{fig:mnist_fmnist}
\end{figure*}

\subsubsection{Lyapunov exponent}

A third diagnostic is the maximal Lyapunov exponent $\lambda$, which 
measures the long-depth average exponential growth rate of a small 
transverse perturbation~\cite{sompolinsky_chaos_1988}. It is the most 
direct probe of the order-to-chaos transition: $\lambda > 0$ signals 
exponential divergence of nearby trajectories (chaotic, 
\texttt{Swish}-dominated phase), $\lambda < 0$ signals exponential 
contraction (ordered, \texttt{Tanh}-dominated phase), and $\lambda = 0$ 
is the critical boundary. Formally, given a base preactivation $z^{(0)}$ 
and a small transverse perturbation $\delta z^{(0)}$, the Lyapunov 
exponent  after $L$ layers is defined by
\begin{equation}
    \|\delta z^{(L)}\| \approx \|\delta z^{(0)}\| e^{\lambda L}.
\end{equation}
At each layer transition $z^{(l+1)} = W^{(l+1)}\sigma(z^{(l)})$, we propagate both the reference state and the perturbation,
\begin{equation}
    \delta z^{(l+1)} = f^{(l)}(z^{(l)} + \delta z^{(l)}) - f^{(l)}(z^{(l)}),
\end{equation}
record the local log-stretch $s_l = \log\!\left(\|\delta z^{(l+1)}\| / \|\delta z^{(l)}\|\right)$,
and immediately renormalize $\delta z^{(l+1)} \leftarrow \varepsilon\, \delta z^{(l+1)} / \|\delta z^{(l+1)}\|$ to prevent numerical overflow or underflow (following the standard Benettin procedure~\cite{benettin_lyapunov_1980}).
The exponent is then estimated by averaging over batch and over the $L_{\mathrm{eff}}$ layers after discarding an initial transient of $l_0$ layers (we take $l_0=5$),
\begin{equation}
    \hat{\lambda} = \frac{1}{L_{\mathrm{eff}}}
    \sum_{l=l_0}^{L-1} \mathbb{E}_{\mathrm{batch}}[s_l].
\end{equation}
At a fixed point of the variance map, $\lambda$ is related to the perpendicular susceptibility
by $\lambda = \tfrac{1}{2}\log \chi_\perp$, so the Lyapunov exponent provides an independent but consistent probe of the same transition. Figure~\figref{fig:susceptibilities}{c} shows a clean zero-crossing at $p \approx p_c$, with $\lambda$ growing continuously from negative to 
positive values as $p$ increases through the critical point, a hallmark of a continuous phase transition between ordered and chaotic phases.

\section{Applications in learning}\label{sec:learning}

\begin{figure*}
    \centering
    \includegraphics[width=\linewidth]{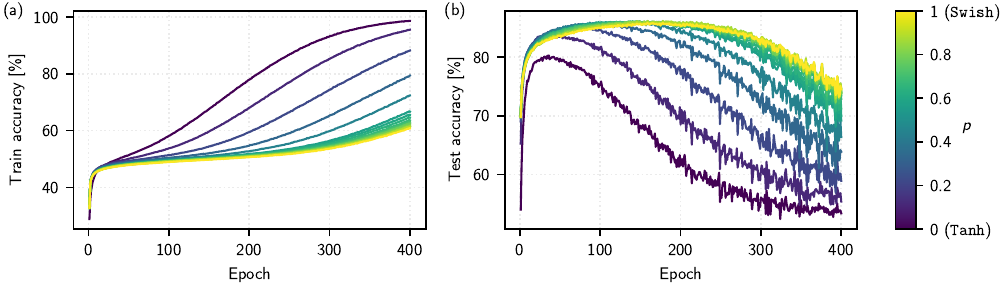}
    \caption{Performance of an overparameterized MLP with a \texttt{Tanh}/\texttt{Swish} activation mixture on the Fashion-MNIST classification task~\cite{xiao_fashion_2017} with 50\% labels corruption.
    (a)~Train accuracy as a function of epoch. All networks eventually memorize the training set, overfitting the corrupted labels, but \texttt{Tanh}-dominated networks (small $p$) get to this undesired point faster.
    (b)~Test accuracy as a function of epoch. All networks shoot up quickly, when learning essentially treats the corrupted labels as noise, but later enter the overfitting regime, where the test accuracy shoots down. We observe non-monotonic behavior with $p$: networks with intermediate values of $p$ (near the supposed critical point) reach their peak test accuracy faster than the \texttt{Swish}-dominated ones, but are more robust to overfitting than the \texttt{Tanh}-dominated ones.
    }\label{fig:over_parameterized}
\end{figure*}

The theoretical framework developed in Sec.~\ref{sec:theory} predicts the critical mixing fraction $p_c$ from the variance map alone, which depends on the network architecture and input statistics but not on the training labels. This has a practically useful consequence: $p_c$ can be estimated from 
forward passes on unlabeled data before any training begins. In practice, one  sweeps $p$ over a coarse grid, feeds a batch of unlabeled inputs through the randomly initialized network, and tracks the depth profile of the variance $K^{(l)}$. The value of $p$ at which the profile is flattest (i.e., closest to 
depth-independent) yields an estimate of $p_c$. This procedure is computationally cheap, requiring only a handful of forward passes at initialization, and entirely label-free. It provides a principled, one-time calibration step that replaces expensive hyperparameter search over activation functions~\cite{bergstra_random_2012, li_hyperband_2017}, and is reminiscent of the mean-field initialization strategies used in, e.g., the weight-agnostic neural network literature~\cite{gaier_weight_2019}.

For both MNIST and Fashion-MNIST, this forward-pass calibration procedure yields $p_c \approx 0.8$, in agreement with the empirically observed critical point $p_c \approx 0.83$ from the variance sweep (Fig.~\ref{fig:variance_plots}). The residual deviation from the small-variance analytical prediction $p_c^{(0)} \approx 0.91$ is accounted for by the finite input variance of the real data, as described by the perturbative correction Eq.~\eqref{eq:pc_corrected}; the agreement between the forward-pass estimate and the variance-sweep estimate confirms that the two procedures locate the same physical transition. The benefit of operating near $p_c$ is expected to become more pronounced in deeper networks, where 
small per-layer deviations from criticality accumulate over depth and exponentially amplify the difference between the critical and off-critical regimes~\cite{schoenholz_deep_2017}. In the following experiments, we deliberately use shallow networks to keep the computational cost tractable 
and to demonstrate that the benefit of the mixture is not contingent on large depth.

\subsection{Non-monotonic test performance and the critical optimum}\label{subsec:mnist_fmnist_performance}

To test whether proximity to criticality translates into learning advantage, we trained a two-hidden-layer MLP of width 64 on two standard image classification benchmarks: the MNIST handwritten digit recognition task~\cite{lecun_gradient_1998} and the more challenging Fashion-MNIST clothing item classification task~\cite{xiao_fashion_2017}.
We trained for 50 epochs using the cross-entropy loss function, with a batch size of 128 and learning rate $10^{-3}$.
For each dataset, we swept the mixing fraction over a dense grid $p \in [0, 1]$ (30 values for MNIST, 10 values for Fashion-MNIST), averaging over 100 random seeds, and recorded the test loss and test accuracy after training.

As shown in Fig.~\ref{fig:mnist_fmnist}, both metrics vary non-monotonically with $p$, with a clear optimum at an intermediate $p$ in both tasks. This 
non-monotonicity is the key signature, as it rules out the possibility that the 
performance gain is simply due to one component dominating, and instead 
points to a genuine benefit of operating near the critical point.
The optimal $p$ is closer to the theoretical value $p_c \approx 0.83$ in MNIST than in Fashion-MNIST.
We attribute the larger deviation in Fashion-MNIST to two factors: the shallowness of the network (two hidden layers is far from the infinite-depth limit in which $p_c$ is defined), and the greater complexity of the Fashion-MNIST task, which makes the optimal initialization more sensitive to non-universal properties of the data and architecture that are not captured by the mean-field theory. Importantly, for Fashion-MNIST, pure 
\texttt{Tanh} ($p = 0$) and pure \texttt{Swish} ($p = 1$) both underperform 
almost all intermediate values of the mixture, demonstrating that the benefit 
is not merely a smooth interpolation between two equally good endpoints but 
reflects a genuine advantage of the mixed, near-critical regime.

\subsection{Quenched disorder as an implicit regularizer}

The statistical mixture has a second distinct practical utility beyond 
initialization quality: it acts as an implicit regularizer in the overparameterized regime.
To understand why, recall that in a homogeneous network---all neurons sharing the same activation function---every neuron is functionally identical at initialization.
In the overparameterized setting, where the number of parameters greatly exceeds the number of training examples, gradient descent can easily coordinate large groups of identical neurons into configurations that memorize spurious input-label 
associations~\cite{zhang_understanding_2017}. This memorization is facilitated by the permutation symmetry of the hidden units: any permutation of neurons within a layer leaves the network function unchanged, so the effective number of independent degrees of freedom available for memorization is not reduced by overparameterization~\cite{dinh_sharp_2017}.

The statistical activation mixture disrupts this by introducing \emph{quenched disorder}~\cite{mezard_spin_1987}: each neuron is permanently assigned either \texttt{Tanh} or \texttt{Swish} at initialization, and this assignment is fixed throughout training.
Because the two activation functions respond to identical weight updates in qualitatively different ways, the permutation symmetry that homogeneous networks exploit for memorization is structurally broken.
Furthermore, memorizing random labels typically requires preactivations to grow large, forming sharp, highly localized decision boundaries~\cite{arpit_closer_2017}.
\texttt{Tanh} neurons saturate when preactivations are large: their gradients vanish and further amplification is suppressed.
Meanwhile, \texttt{Swish} neurons maintain gradient flow for the large-scale features that support generalization.
Together, these two mechanisms bias optimization toward flatter, more generalizable minima~\cite{hochreiter_flat_1997, keskar_large_2017}, without any explicit regularization term in the loss.

To test this mechanism, we trained an overparameterized MLP with four hidden layers of widths $\left\{1024,1024,512,512\right\}$ on Fashion-MNIST~\cite{xiao_fashion_2017} under severe label corruption: 50\% of training labels were randomly reassigned, forcing the network to choose between learning genuine image structure and memorizing noise. The test set remained uncorrupted throughout. 
To facilitate a direct comparison of different values of $p$, we used the plain stochastic gradient descent optimizer. The other hyperparameters are the same as in Sec.~\ref{subsec:mnist_fmnist_performance}, except we trained for 400 epochs.

As shown in Fig.~\ref{fig:over_parameterized}, all networks eventually overfit the corrupted labels, but the test accuracy trajectory strongly depends on $p$.
\texttt{Tanh}-dominated networks ($p \to 0$) memorize the training set fastest and suffer the most severe test accuracy degradation as corrupted labels are absorbed. \texttt{Swish}-dominated networks ($p \to 1$) are slower to memorize but still eventually overfit. Networks near the critical mixing fraction $p_c$ offer the best of both regimes: they reach their peak test accuracy faster than \texttt{Swish}-dominated networks---because the Swish component keeps gradient flow alive for genuine 
structure---and sustain it longer than \texttt{Tanh}-dominated ones---because 
the Tanh component suppresses the large preactivations needed for 
memorization. This demonstrates that quenched activation disorder suppresses 
memorization while preserving the capacity to learn genuine structure.

\section{Outlook}\label{sec:conclusion}

The central result of this work is that the universality classes of deep neural networks, conventionally treated as discrete labels distinguished by the qualitative structure of the variance recursion near $K^{\star}=0$~\cite{poole_exponential_2016,schoenholz_deep_2017,roberts_principles_2022}, are in fact connected by a continuous family of \emph{statistical activation mixtures}.
Drawing each neuron's activation independently from a Bernoulli distribution over two functions $\{\sigma_1,\sigma_2\}$ reduces, by self-averaging at infinite width, to a linear interpolation of the kernel maps, Eq.~\eqref{eq:g_mix}.
The mixing fraction $p$ thus becomes an analytically transparent control parameter that continuously deforms the variance map between the two pure limits, with a closed-form critical point $p_c$ given by Eq.~\eqref{eq:pc_general}.
When $\sigma_1$ and $\sigma_2$ belong to opposing classes, such as \texttt{Tanh} (stable $K^{\star}=0$) and \texttt{Swish} (half-stable), the transition separates a variance-collapsing phase from a variance-inflating one (both with power-law behavior), and is empirically diagnosed by three observables: variance propagation, susceptibilities and the Lyapunov exponent. The transition is not merely an initialization artifact: it has direct 
consequences for learning, manifesting as non-monotonic test performance with an optimum near $p_c$, and as quenched-disorder regularization that suppresses memorization under label corruption. Together, these results establish statistical activation mixtures as a controlled, analytically tractable tool for navigating the phase diagram of deep network universality classes.

The statistical (``incoherent'') and deterministic (``coherent'') constructions are two natural ways to combine activations, and their analytical inequivalence is not merely a technical distinction; it reflects a fundamental difference in the physical structure of the problem.
The coherent sum mixes at the level of each neuron's response, so cross-correlation kernels $\tilde{g}(K)=\langle\sigma_1\sigma_2\rangle_K$ enter the recursion and the dependence on $p$ is nonlinear.
The statistical mixture instead mixes at the level of the ensemble, with each neuron quenched to one activation at initialization~\cite{bahri_statistical_2020,pei_statistical_2025}; self-averaging renders the kernel map linear in $p$, giving closed-form expressions for the critical mixing and the universality classes on either side.
The same quenched heterogeneity underlies the regularization effect we observed under label corruption, by breaking the permutation symmetry that homogeneous networks exploit to memorize noise~\cite{zhang_understanding_2017}.  This connection between the 
analytical tractability of the mixture and its practical regularization properties is not coincidental: both stem from the same structural feature, i.e. the independence of each neuron's activation assignment.

From a practical standpoint, the mixture yields a label-free, forward-pass-only protocol for selecting an activation architecture.
Because $p_c$ is fixed by the input statistics and the architecture alone, it can be estimated before any training by locating the mixing fraction at which $K^{(l)}$ is flattest in depth, or analytically via Eq.~\eqref{eq:pc_general} with a perturbative correction for finite input variance, Eq.~\eqref{eq:pc_corrected}.
This replaces costly hyperparameter searches over activation functions with a one-shot calibration costing only a handful of forward passes at initialization. The strategy should scale favorably to larger models~\cite{kaplan_scaling_2020}; as networks grow deeper, the signal-propagation benefits of criticality become more pronounced, and the 
cost of the forward-pass calibration grows only linearly with depth while the 
cost of training grows much faster.

Importantly, the critical mixture realizes the scale-invariant propagation of \texttt{ReLU} using components that are everywhere smooth and infinitely differentiable.
\texttt{ReLU}'s non-smoothness, its vanishing second derivative away from the cusp, and its ill-defined Hessian at $z=0$, make it ill-suited to any method that probes curvature, like natural-gradient and Hessian-based optimizers ~\cite{amari_natural_1998, martens_new_2020}, neural tangent kernel analyses at finite width~\cite{jacot_neural_2018}, and geometry-aware variational autoencoders~\cite{kim_gamma_2024}.
It also makes it unusable in architectures where smoothness is a physical requirement rather than a convenience, like physics-informed neural networks that solve partial differential equations by differentiating through the 
network~\cite{raissi_physics-informed_2019}, and neural-network quantum states 
whose variational energy involves derivatives of the wavefunction~\cite{carleo_solving_2017,lange_architectures_2024}.
The standard remedies, \texttt{GELU}, \texttt{Swish}, and \texttt{ELU}~\cite{hendrycks_gaussian_2016,ramachandran_searching_2017,clevert_fast_2016}, introduce a length scale and place the network in the half-stable class, where the variance is driven to a non-zero fixed point and deep propagation is compromised.
A \texttt{Tanh}/\texttt{Swish} mixture tuned to $p_c$ delivers both properties simultaneously: approximate statistical scale invariance for robust depth scaling, and $C^\infty$ smoothness for reliable higher-order differentiation. We view this as the most immediately actionable consequence of the theory for practitioners working in these 
domains.

The transition described in this work belongs to a broader family of 
order-to-chaos transitions driven by competing local operations with opposing tendencies, a mathematical structure that has emerged independently in several areas of physics. As noted previously, the closest parallel is with MIPTs in monitored quantum 
circuits~\cite{li_quantum_2018, nahum_quantum_2017, 
skinner_measurement_2019, fisher_random_2023}. In that setting, 
entangling unitary gates compete against disentangling projective 
measurements; tuning the relative rate $p$ of measurements drives a 
continuous transition between a volume-law entangled phase and an area-law 
phase, diagnosed by the entanglement entropy and its analogs. The field-theoretic machinery developed 
for these transitions~\cite{bao_theory_2020, 
jian_measurement_2020,nahum_quantum_2017,li_conformal_2021} may find direct application in the deeper analysis of activation-mixture criticality in the future. 

Several extensions of the present framework follow naturally, and we 
highlight those we consider most promising.
The Bernoulli ensemble is the simplest nontrivial ${\cal P}(\sigma)$. More general discrete or continuous distributions over activation parameters admit the same mean-field treatment via Eq.~\eqref{eq:g_mix} and may expose richer critical manifolds, including multicritical points where three or more universality classes meet. The order parameter $a_1^{(\rm mix)}(p)$ generalizes straightforwardly to a function on the space of distributions $\mathcal{P}(\sigma)$, and the condition $a_1^{(\rm mix)} = 0$ defines a critical hypersurface in this space. Mapping this hypersurface is a natural next step. The quenched assignment studied here can be extended to an \emph{annealed} rule in which each neuron independently redraws its activation on every forward pass, in the spirit of stochastic-activation schemes recently explored for inference-time diversity~\cite{lomeli_stochastic_2025}; we expect this to affect the regularization behavior without altering the mean-field location of $p_c$. Quantifying this difference, both theoretically and empirically, would clarify the relative contributions of the critical initialization and the quenched heterogeneity to the observed learning benefits.

Extending the present analysis to architectures with structured nonlinearities, e.g. convolutional layers~\cite{xiao_dynamical_2018}, attention mechanisms~\cite{vaswani_attention_2017}, and layer normalization~\cite{ba_layer_2016}, requires generalizing the kernel 
recursion to account for the spatial structure of the activations and the 
normalization-induced coupling between neurons. The universality-class structure associated with the analogous transitions is not well understood, and the activation-mixture framework could provide a new handle on initialization and signal propagation in these architectures.

The mean-field theory gives a closed-form $p_c$ and predicts $\nu = 1$, which 
we confirm numerically via finite-size scaling of the depth profile (Fig.~\ref{fig:fss}). However, the mean-field exponent is generically modified by fluctuations beyond the infinite-width 
limit~\cite{cardy_scaling_1996}. At finite width $N$, both $p$ and $L$ enter the scaling theory, and the relevant scaling variable is expected to become $(p - p_c) L^{1/\nu} f(L/N^\alpha)$ for some crossover exponent $\alpha$ that encodes the finite-width corrections. Extracting these exponents numerically and comparing them to predictions from a putative field theory of the transition, including drawing on the MIPT analogy discussed above, would establish whether the activation-mixture transition defines a new universality class, or falls into a known class of order-to-chaos transitions.

\section*{Acknowledgment}
We are grateful to M.~Barkeshli, C.~Myers, Z.~Ringel, J.~P.~Sethna, and J.~Tahmassebpour for valuable comments on the manuscript.
O.L. acknowledges support from the Bethe-KIC postdoctoral fellowship at Cornell University. D.C. and O.L. are supported in part by a grant from the Department of Energy (DE-SC0026112) under the Early Career Research Program to D.C.
The authors acknowledge the use of large language models (Claude, ChatGPT, and Gemini) for code writing and polishing the manuscript.

\emph{Data Availability:} The codebase used in this work is publicly available at \url{https://doi.org/10.5281/zenodo.19683547}.

\appendix

\section{Mixtures containing \NoCaseChange{\texttt{ReLU}}: \\
absence of a phase transition}
\label{app:relu}

Here we show a basic case where the statistical mixture does not yield a phase transition, to illustrate the importance of both components having $g_2 \neq 0$ for the existence of a transition at $p_c < 1$.
Consider a simple test case by mixing \texttt{ReLU} and another activation from one of the other universality classes, such as \texttt{Tanh}.
This case does not yield a useful construction for our purposes, since \texttt{ReLU} is already scale-invariant, so mixing it 
with a smooth activation does not resolve the smoothness problem. However, it provides a clean illustration of why the phase transition of 
Eq.~\eqref{eq:pc_general} requires both components to have $g_2 \neq 0$, 
i.e., both must belong to non-scale-invariant classes.

\texttt{ReLU}'s scale invariance, $\sigma(\alpha z) = \alpha \sigma(z)$, 
forces its kernel function to be exactly linear: $g^{(\rm ReLU)}(K) \propto 
K$, so $g_2^{(\rm ReLU)} = 0$ identically. This is not an approximate 
statement that holds in the small-$K$ limit; it is an exact algebraic 
consequence of scale invariance that holds for all $K$. Any activation from 
the stable class (e.g., \texttt{Tanh}) or the half-stable class (e.g., 
\texttt{Swish}) has $g_2 \neq 0$. Substituting $g_2^{(\sigma_1)} = 
g_2^{(\rm ReLU)} = 0$ into Eq.~\eqref{eq:pc_general} immediately gives
\begin{eqnarray}
    p_c = \frac{g_2^{(\rm other)}}{g_2^{(\rm other)} - 0} = 1,
\end{eqnarray}
for any choice of the second component. There is therefore no transition at any $p<1$. The scale-invariant fixed point of \texttt{ReLU} is 
structurally unstable to any admixture of a non-scale-invariant activation. 
Physically, even an infinitesimal fraction $(1-p)$ of \texttt{Tanh} neurons 
introduces a nonzero $g_2^{(\rm mix)}$, which immediately places the network 
in a non-marginal universality class.

For the specific case of a \texttt{ReLU}/\texttt{Tanh} mixture, $g_2^{(\rm Tanh)} = -2 < 0$, so $a_1^{(\rm mix)}(p) < 0$ for all $p < 1$. The \texttt{Tanh} component dominates the stability, and the network is driven into the $K^\star = 0$ stable class regardless of how small the \texttt{Tanh} fraction is. Scale invariance is, in this precise sense, non-generic: it requires $g_2 = 0$ exactly, a condition that cannot be maintained under perturbations that introduce a finite length scale.

\begin{figure}
    \centering
    \includegraphics[width=\linewidth]{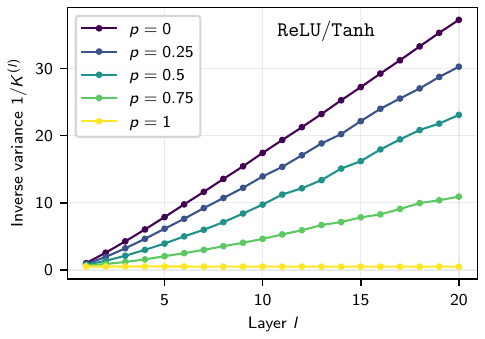}
    \caption{Inverse variance $K^{(l)}$ vs.\ depth $l$ for a \texttt{ReLU}/\texttt{Tanh} activation mixture, for several values of the mixing fraction $p$.
    Because \texttt{ReLU} is scale-invariant ($a_1=0$), the linear term in the variance map always vanishes and $K^{(l)}\sim 1/l$ for any $p<1$. Consequently, no phase transition exists: the network always remains in the \texttt{Tanh}-dominated regime ($K^{\star}=0$ class).}
    \label{fig:relu_tanh_variance}
\end{figure}

We verify this numerically in Fig.~\ref{fig:relu_tanh_variance}, which shows 
the inverse variance $1/K^{(l)}$ as a function of depth $l$ for randomly 
initialized MLPs with a \texttt{ReLU}/\texttt{Tanh} mixture at several 
values of $p$. For all $p < 1$, the inverse variance grows linearly with 
depth ($K^{(l)} \sim 1/l$), the algebraic decay characteristic of the 
$K^\star = 0$ stable class, without any signature of a transition. This 
confirms that a phase transition between the scale-invariant and stable 
classes requires non-generic fine-tuning $p \to 1$, i.e., the complete 
elimination of the \texttt{Tanh} component. The practical implication is 
that smooth scale-invariant propagation cannot be achieved by diluting 
\texttt{ReLU} with a smooth activation; one must instead work entirely 
within the smooth classes and engineer a transition between them, which is 
the strategy pursued in the main text.

\section{Additional data for variance propagation}\label{app:variance_plots}

\begin{figure}
    \centering
    \includegraphics[width=\linewidth]{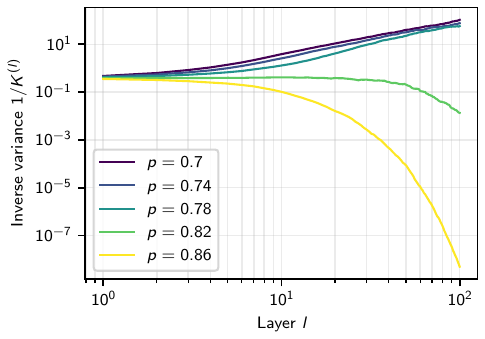}
    \caption{Variance propagation for a \texttt{Tanh}/\texttt{Swish} mixture with $L=100$ layers, plotted on a log-log scale. The variance decays algebraically with depth when $p < p_c$ and grows algebraically when $p > p_c$.}
    \label{fig:variance_deep_loglog}
\end{figure}

\begin{figure}
    \centering
    \includegraphics[width=\linewidth]{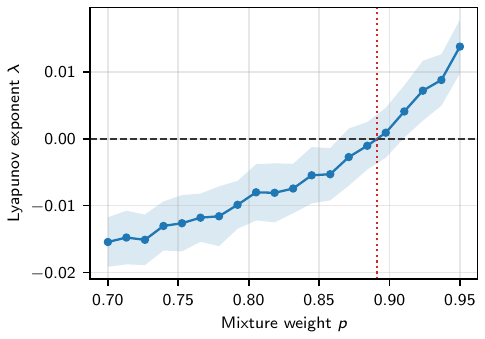}
    \caption{Lyapunov exponent for a \texttt{Tanh}/\texttt{Swish} mixture with a smaller initial variance $K_0=0.05$. The critical point, which is where the Lyapunov exponent crosses zero, shifts to $p_c \approx 0.89$. This is closer to the small-variance prediction $p_c^{(0)} \approx 0.91$ than the critical point observed in Fig.~\figref{fig:susceptibilities}{c} of the main text, which is $p_c \approx 0.83$. The shift is explained by the perturbative correction for finite input variance, Eq.~\eqref{eq:pc_corrected}.}
    \label{fig:lyapunov_small_variance}
\end{figure}

Here we show additional simulation results for the variance propagation, which complement the main text. 
Figure~\ref{fig:variance_deep_loglog} shows the variance propagation for a \texttt{Tanh}/\texttt{Swish} mixture for $L=100$ layers (other parameters are the same as in Fig.~\ref{fig:variance_plots} of the main text). The power-law behavior is more clearly visible on the log-log scale, with the variance decaying algebraically with depth when $p < p_c$ and growing algebraically when $p > p_c$. Deviations from the power-law behavior appear at large depth.

Figure~\ref{fig:lyapunov_small_variance} shows the Lyapunov exponent for a \texttt{Tanh}/\texttt{Swish} mixture with a smaller initial variance $K_0=0.05$, which pushes the critical point upward toward the small-variance prediction $p_c^{(0)} \approx 0.91$. Since the variance is small, more random seeds are needed for numerical stability and convergence: the results shown in Fig.~\ref{fig:lyapunov_small_variance} are averaged over 100 random seeds, whereas the results in the main text, specifically Fig.~\figref{fig:susceptibilities}{c}, reach convergence at about 10 random seeds.

\bibliography{library}

\end{document}